\documentclass[galaxies,article,submit,moreauthors,10pt,a4paper]{mdpi} 

\firstpage{1} 
\articlenumber{29}
\doinum{10.3390/galaxies6010029}
\pubvolume{6}
\pubyear{2018}
\copyrightyear{2018}
\history{Received: 8 February 2018; Accepted: 20 February 2018; Published: 28 February 2018}

\usepackage{amssymb}

 \theoremstyle{mdpi}
 \newcounter{thm}
 \newcounter{ex}
 \newcounter{re}
 \theoremstyle{mdpidefinition}

\Title{Unveiling the Origin of the Fermi Bubbles}

\Author{H.-Y. Karen Yang $^{1,2}$, Mateusz Ruszkowski $^{3}$ and Ellen G. Zweibel $^{4}$}
\AuthorNames{H.-Y. K. Yang, M. Ruszkowski and E. G. Zweibel}

\address{%
$^{1}$ \quad Department of Astronomy, University of Maryland, College Park, {MD} 20742,  USA; hsyang@astro.umd.edu\\
$^{2}$ \quad Joint Space-Science Institute, College Park, {MD} 20742,  USA; hsyang@astro.umd.edu\\
$^{3}$ \quad Department of Astronomy, University of Michigan, Ann Arbor, {MI} 48109, USA; mateuszr@umich.edu\\
$^{4}$ \quad Departments of Astronomy \& Physics, University of Wisonsin-Madison, Madison, {WI} 53706, USA; zweibel@astro.wisc.edu}

\abstract{The {\it Fermi} bubbles, two giant structures above and below the Galactic center (GC), are among the most important discoveries of the {\it Fermi Gamma-ray Space Telescope}. Studying their physical origin has been providing valuable insights into cosmic-ray transport, the Galactic magnetic field, and past activity at the GC in the Milky Way galaxy. Despite their importance, the formation mechanism of the bubbles is still elusive. Over the past few years there have been numerous efforts, both observational and theoretical, to uncover the nature of the bubbles. In this article, we present an overview of the current status of our understanding of the bubbles' origin, and discuss possible future directions that will help to distinguish different scenarios of bubble formation.}

\keyword{gamma rays; Galaxy; cosmic rays.}

\begin{document}




\section{Introduction}

The {\it Fermi} bubbles are two giant bubbles extending $\sim 50^\circ$ above and below the Galactic Center (GC) revealed by the {\it Fermi Gamma-ray Space Telescope} \cite{Su10}. They are one of the three elephants in the gamma-ray sky (among Loop I and the Galactic Center Excess): the solid angle of the bubbles is about 1 sr, which is roughly that of an elephant standing in a room. They are "elephant in the room" also because of their mysterious physical origin. Their symmetry about the GC suggests that they originate from powerful energy injections from the GC, possibly related to nuclear star formation (NSF) or active galactic nucleus (AGN) activity. Because of the proximity, there are ample data from multi-messenger observations in the vicinity of the {\it Fermi} bubbles, ranging from radio, microwave, X-ray, gamma-ray, to neutrinos. These spatially resolved, multi-messenger data offers unprecedented opportunities to learn about cosmic-ray (CR) transport, the Galactic magnetic field, and past activity of the GC in our Milky Way galaxy. 

Despite their importance, the physical origin of the {\it Fermi} bubbles is still elusive. Various theoretical models have been proposed to explain the formation of the bubbles. However, it has been challenging for any model to match all the observational data simultaneously, which is critical before any model can claim success. A primary goal of this review is to discuss the strengths and weaknesses of each model using the observational data as rubrics, with the hope of providing some guidance for future improvements. 

In \S~\ref{sec:obs}, we summarize the observable properties of the {\it Fermi} bubbles, which are important constraints on the theoretical models. In \S~\ref{sec:theory} we present an overview of theoretical models proposed to date. We first discuss some general considerations that go into the model construction and divide the models into three categories (\S~\ref{sec:general}), and then summarize important findings of each category (\S~\ref{sec:hadronic}-\S~\ref{sec:insitu}). Finally, we identify possible directions of future research that could help disentangle the different scenarios of bubble formation. 

\section{Observable Properties of the Bubbles}
\label{sec:obs}

In this section we give an overview of the existing multi-messenger observational data in the vicinity of the {\it Fermi} bubbles, which form a set of constraints that the theoretical models of bubble formation must satisfy. 

The gamma-ray bubbles are observed by {\it Fermi} between 1 GeV and $\sim 100$ GeV with a hard spectrum of spectral index of approximately -2. The total gamma-ray power for both bubbles is $\sim 2.5 \times 10^{40}$ GeV s$^{-1}$ or $\sim 4.0 \times 10^{37}$ erg s$^{-1}$. The gamma-ray bubbles have many unique characteristics, including smooth surface, sharp edges, and almost flat intensity distribution \cite{Su10}. More recent data has confirmed these basic features, through providing better constraints on the sharp edges, substructures in the surface brightness distribution, and spatial variation of the spectrum \cite{Ackermann14, Narayanan17}. Specifically, the previously identified "cocoon" structure on top of the flat surface brightness distribution by \cite{Su12} is confirmed, albeit the second jet feature has become insignificant. A high-energy cutoff at $\sim 110$ GeV is revealed in the gamma-ray spectrum. Although there are some hints of spectral variation near the bubble edges \cite{Yang14, Keshet17a}, the gamma-ray spectrum is remarkably latitude independent, including the amplitude (a.k.a.\ the flat intensity), overall shape, and the high-energy cutoff energy. To reproduce the spatially uniform spectrum of the bubbles is nontrivial as it requires correct representation of the underlying CR distribution both spatially and spectrally. 

The gamma-ray bubbles are in fact counterparts of the microwave haze first detected by the {\it Wilkinson Microwave Anisotropy Probe} \cite{Finkbeiner04} and later confirmed by the {\it Planck} satellite \cite{PlanckHaze}. Although the haze was originally speculated to be a signature of dark matter annihilation, this hypothesis is disfavored because of the sharp edges of the gamma-ray bubbles revealed by {\it Fermi}. The gamma-ray bubbles and the microwave haze are spatially correlated and both exhibit very hard spectra, suggesting a common origin. In contrast to the uniform intensity distribution in the gamma-ray band, the haze dims away from the GC and there appears to be a suppression beyond a Galactic latitude of $\sim 35^\circ$ \cite{Dobler12a}. As we will discuss in \S~\ref{sec:theory}, it has been a challenge for some of the models to reproduce both the gamma-ray and microwave emissions simultaneously.

Radio polarization data  by the {\it S-band Polarization All Sky Survey (S-PASS)} at 2.3 GHz is also available in the vicinity of the {\it Fermi} bubbles \cite{Carretti13}. Two highly polarized lobes that have similar shapes to the gamma-ray bubbles are identified, though extending to $\sim\pm$ 60$^\circ$. Instead of the spatially uniform hard spectrum of the gamma-ray bubbles, the spectrum of the polarized lobes is softer toward higher Galactic latitudes. The high degree of polarization suggests regularity of the magnetic field in the surroundings of the bubbles. 

In X-rays, outflows from elevated past GC activity was also suggested prior to the detection of the gamma-ray bubbles \cite{BlandHawthorn03}. The location of the gamma-ray bubbles is X-ray dim \cite{Snowden97}, suggesting the bubble interior is underdense. Arc features in the {\it ROSAT} X-ray map are coincident with the edges of the gamma-ray bubbles, hinting at the presence of gas compressed by shocks \cite{Su10}. More recently, attempts have been made to compile X-ray and UV data for multiple sightlines to probe the thermal and the kinematic structures of the Galactic halo \cite{Kataoka13, Miller13, Fang14, Kataoka15, Fox15, Miller16, Bordoloi17, Sarkar17}. Although these studies have brought valuable insights into the origin of the bubbles, there is a mix of results: both slow (several hundreds of km s$^{-1}$) and fast (> 1000 km s$^{-1}$) outflows have been inferred. The discrepancies could be due to a number of factors: (1) the structure of the Galactic halo is complex \cite{Kataoka15}; (2) confusion due to foreground or background projections into the lines of sight; (3) assumptions about the outflow geometry and injection patterns \cite{Fox15, Miller16, Sarkar17}; (4) non-negligible timescales for electron-proton equilibration not accounted for \cite{Keshet17b}; and (5) gas probed by X-ray and UV is in fact of different temperatures and phases. Because of the above reasons, care should be taken when one interprets the results.  

There are constraints on the bubble emission at energies higher than the {\it Fermi} band. For instance, the {\it High Altitude Water Cherenkov (HAWC)} observatory has recently derived upper limits for the bubbles in the TeV range, showing suppression of bubble emission beyond 100 TeV \cite{HAWC17}. The exact cutoff energy of the bubble spectrum will be better constrained as {\it Fermi, HAWC}, and other TeV gamma-ray observatories collect more data (see \S~\ref{sec:future}). The presence of PeV neutrinos could also be used to infer the existence of high-energy CR protons (CRp) from the bubbles. Although {\it IceCube} has detected neutrinos with incoming directions coincident with the bubbles, estimating the neutrino background has been a difficult task due to the low number counts and so far it is still unclear whether there are neutrinos associated with the bubbles themselves after background subtraction \cite{Razzaque13, Ahlers14, Fang17, Sherf17}.

\section{Overview of Theoretical Models}
\label{sec:theory}

In this section, we review the important findings of the theoretical models that have been proposed to explain the origin of the {\it Fermi} bubbles. We first lay out some general considerations that go into the model construction and divide the models into three categories (\S~\ref{sec:general}). We then summarize the status of each category and discuss their strengths and weaknesses (\S~\ref{sec:hadronic}-\S~\ref{sec:insitu}). Note that our aim is to briefly summarize the current status and to depict the general trend instead of scrutinizing each theoretical model. Therefore we refer the readers to the individual references for details that are not included in this article. 

\subsection{General Considerations}
\label{sec:general}

In terms of understanding the physical origin of the {\it Fermi} bubbles, three major questions need to be addressed. First, what is the emission mechanism? The bubbles can either be hadronic, where the gamma rays are produced by inelastic collisions between CRp and the thermal nuclei via decay of neutral pions, or leptonic, where the gamma rays are generated by inverse-Compton (IC) scattering of the interstellar radiation field (ISRF) by CR electrons (CRe). Second, what activity at the GC triggered the event -- are the bubble associated with NSF or AGN activity? Third, where are the CRs accelerated? They could either be accelerated at the GC and transported to the surface of the bubbles, or accelerated in-situ by shocks or turbulence.  

Note however that not all combinations of the above three considerations would make a successful model because of constraints given by the hard spectrum of the observed bubbles. For instance, if the bubbles are leptonic, the synchrotron and IC cooling time of high-energy ($\sim$ TeV) CRe gives a very stringent age limit on the bubbles to be within a few million years old. Therefore, fast AGN jets with velocities on the order of thousands of km s$^{-1}$ are required to maintain the hard spectrum if the CRs are transported from the GC to large heights to the disk. If instead one wishes to build a model based on transport of CRs from the GC by NSF or AGN winds of typical speeds of hundreds of km s$^{-1}$, then the bubbles have to be hadronic, otherwise the CRe would have cooled. Alternately, one can bypass the age constraints completely by invoking {\textit{in-situ}} acceleration of CRs near the bubble surface by shocks or turbulence. To this end, any successful model has to fall into one of the following three categories: (1) hadronic wind models, (2) leptonic jet models, and (3) {\textit{in-situ}} acceleration models. We discuss each category in the following sections. We note that, while it would be instructive to estimate the total energy required to inflate the bubbles for each type of model, the answers vary over a wide range depending on the model assumptions. Estimates of total bubble energies from observations \cite{Su10, Miller16} are also highly dependent on the values assumed for physical parameters. Therefore, useful constraints from energetics are not available at this time.

\subsection{Hadronic Wind Models}
\label{sec:hadronic}

In these models, CRp are accelerated at the GC and transported by NSF \cite{Crocker11, Crocker15} or AGN \cite{Mou14, Mou15} winds. Since the winds have typical velocities of hundreds of km s$^{-1}$, the bubbles are formed on timescales of $\gtrsim$ 10 Myr. To reproduce the lobular shape of the observed bubbles, the quasi-spherical winds would need to be collimated, e.g., by the central molecular zone \cite{Zubovas12, Mou14}. 

In hadronic models that invoke NSF winds, because of the low gas density within the bubbles and the resulting long timescales of the hadronic collisions, the bubbles have to be formed on a timescale of multiple billion years in order to match with the observed gamma-ray luminosity given typical star formation rate within the Galaxy \cite{Crocker11}. Since such a timescale is much longer than typical timescales of other relevant physical processes associated with galactic winds, more recent models have invoked gas condensation from thermal instabilities in order to shorten the formation time to $\gtrsim$ 100 Myr \cite{Crocker15}. 

Outflows from the central AGN have also been proposed as a possible mechanism for bubble formation \cite{Mou14, Mou15}, motivated by some observational evidence for elevated past activity of the Sgr A* \cite{Totani06}. Using three-dimensional (3D) simulations of AGN winds, \cite{Mou14} showed that the {\it Fermi} bubbles could be inflated by an active phase of Sgr A* which started about 10 million years ago and was quenched no more than 0.2 million years ago. The expansion velocity is in agreement with the gentler outflows inferred by earlier X-ray studies \cite{Kataoka13} (despite the caveats discussed in \S~\ref{sec:obs}). The 3D magnetohydrodynamic (MHD) simulations of \cite{Mou15} also showed that the gamma-ray characteristics of the observed bubbles can be reproduced by a combination of primary CRp and secondary CRe. 

One of the major challenges for the hadronic models is to reproduce the microwave haze: the predicted microwave emission from the secondary electrons and positrons is too low compared to the observed one, and the spectrum is too soft \cite{Ackermann14, Cheng15a}. Therefore, another population of primary CRe is required in order to match the observed haze emission. In the model of \cite{Crocker15}, the problem is overcome by introducing a giant reverse shock $\sim$ 2 kpc away from the Galactic disk, which could produce freshly accelerated CRe that are responsible for the microwave haze. In addition, the CRe age as they travel to large heights away from the disk, producing the lower energy CRe necessary to generate the polarized lobes as observed by {\it S-PASS}. It is unclear, though, whether this model would be consistent with the flat gamma-ray intensity distribution (since the condensation preferentially occurs at higher latitudes) and the hard microwave spectrum after line-of-sight (LOS) projections (since any sightline would pass through different heights of the bubbles and thus different ages of the CRe). Future 3D simulations of this scenario will be instrumental for addressing these questions.    

In short summary, the overall properties of the observed gamma-ray bubbles could be successfully explained by hadronic interactions between the NSF or AGN winds and the interstellar medium. However, purely hadronic models fail to reproduce the microwave haze and hence another population of primary CRe is needed. A giant reverse shock could be a plausible source of primary CRe that are responsible for the microwave and radio polarization data \cite{Crocker15}, though it remains to be seen whether the model holds after taking into account the effect of LOS projections. In addition, the high-energy cutoff in the observed gamma-ray spectrum still needs to be explained.  

\subsection{Leptonic Jet Models}
\label{sec:leptonic}

The {\it Fermi} bubbles, if produced by CRe accelerated from the GC, must be inflated within a few million years, before the CRe cool. Two-dimensional hydrodynamic simulations of \cite{Guo12a} demonstrated the plausibility of bubble formation by fast AGN jets consistent with the age constraint. The simulated bubble surface is rippled due to fluid instabilities, however, and hence viscosity is proposed as one possible mechanism for producing the observed smooth surface \cite{Guo12b}. 

The leptonic jet scenario is further explored in detail by 3D MHD simulations of \cite{Yang12, Barkov14}. By incorporating additional observational constraints on gas temperature within the bubbles from X-ray absorption lines \cite{Miller13}, Yang et al.\ \cite{Yang12} found that the age of the bubbles can be reduced to $\gtrsim$ 1 Myr, which naturally yields the smooth surface since the formation time is shorter than the growth time of the instabilities. The sharp edges of the bubbles could be due to anisotropic CR diffusion along magnetic field lines that drape around the bubble surface during bubble expansion \cite{Yang12}. Taking into account LOS projections, the gamma-ray, X-ray, microwave, and polarization properties of the simulated bubbles are in agreements with the observed ones \cite{Yang12, Yang13}. Importantly, the microwave haze and the gamma-ray bubbles can be reproduced by the same population of primary CRe. More recently, it was further shown that the high energy cutoff at $\sim$ 110 GeV could be a signature of synchrotron and IC cooling of CRe near the GC when the jets were first launched \cite{Yang17}. 
In their model, the spatially uniform spectrum can be explained since the dynamical time of the jets is shorter than other relevant cooling times after the CRs suffer from the initial cooling losses near the GC. In addition, adiabatic compression by the jets during the active phase causes slightly enhanced CR spatial and energy distribution at higher latitudes, which compensates for the gradient of the ISRF away from the Galactic plane. For these reasons, this model is able to reproduce the spatial uniformity of observed gamma-ray spectrum, including the amplitude, shape, and high energy cutoff.   

In brief, the leptonic jet scenario is in agreement with the observed properties of the {\it Fermi} bubbles. The pros of this model include the fact that the microwave haze emission can be generated simultaneously by the same CRe that produce the gamma-ray bubbles, and that the high energy cutoff in the gamma-ray spectrum is more easily explained. One disadvantage of this model is that it requires very high accretion rates ($\gtrsim 10\%$ of the Eddington accretion rate) during the active phase of the Sgr A*, though there is tentative evidence for such flare activity \cite{BlandHawthorn13}. Moreover, it remains to be seen whether the fast outflow velocities predicted by this model would be consistent with those inferred from X-ray and UV studies of the thermal and kinematic structures of the Galactic halo.  

\subsection{{\textit{In-situ}} Acceleration Models}
\label{sec:insitu}

In the {\textit{in-situ}} acceleration models, CRs are assumed to be accelerated by shocks \cite{Cheng11, Fujita13, Lacki14, Fujita14, Cheng15b, Keshet17a} or turbulence \cite{Mertsch11, Cheng14, Sasaki15} within the bubbles, preferentially near the bubble edges as required by the flat gamma-ray intensity distribution. The gamma rays are generated by freshly accelerated CRs near their production site, naturally producing the sharp edges and satisfying the age constraints given by the hard spectrum. The shocks and turbulence could be triggered by NSF \cite{Lacki14, Sarkar15}, accretion \cite{Zubovas11, Zubovas12} or tidal disruption events (TDEs) \cite{Cheng11, Cheng15b} of the AGN, or an un-specified event \cite{Mertsch11, Fujita13, Fujita14, Cheng14, Sasaki15, Keshet17a}. The emission mechanism could be leptonic \cite{Cheng11, Mertsch11, Cheng14, Cheng15b, Sarkar15, Sasaki15}, hadronic \cite{Fujita13}, or a combination of both \cite{Lacki14, Fujita14}.  

One of the challenges of the {\textit{in-situ}} acceleration models is to reproduce the flat surface brightness distribution of the observed bubbles. The shocks produced by periodic TDEs are expected to yield a constant gamma-ray volume emissivity distribution \cite{Cheng11}, rather than the constant projected surface brightness distribution as observed. In the simplest models, the CRs are concentrated in a thin shell near the bubble edges, and therefore the projected intensity profile tends to be edge brightened \cite{Mertsch11, Fujita13, Sasaki15}. In order to broaden the CR distribution, more recent models have considered efficient escape of CRe \cite{Sasaki15} or hadronic emission since CRp are more long-lived \cite{Fujita13}. However, these modifications have added complications in the spectra, making it a nontrivial task to reproduce the microwave haze emission and the spatially uniform gamma-ray spectrum \cite{Fujita14, Sasaki15}.

Another limitation of the {\textit{in-situ}} acceleration models is that, partly because the injection event is often uncertain, simplified assumptions have to be made, such as assumptions about the CR distribution \cite{Sarkar15} and spherical symmetry of the bubbles. Because the observed bubbles are large and elongated, and the properties within the Galactic halo (e.g., gas density, magnetic field strength, photon density of the ISRF) vary with location, it is crucial to take into account the effects of LOS projections. Realistic 3D simulations including CRs will be needed to aid accurate interpretation of the data. 

To summarize, the advantages of the {\textit{in-situ}} acceleration models include that they are free from the age constraints and that the sharp edges are more easily produced. However, more realistic 3D simulations including LOS projections are required to assess whether the models are consistent with the spatial and spectral properties of the multi-wavelength data.     



\section{Concluding Remarks and Future Prospects}
\label{sec:future}

Since the discovery of the {\it Fermi} bubbles, substantial progress has been made to uncover their physical origin, both observationally and theoretically. Although the formation mechanisms are still debated, valuable insights have been gained thanks to the collegial efforts of the community. In particular, we have learned that purely hadronic models fail to reproduce the microwave haze and require another population of primary CRe in order to match the haze emission. The simplest {\textit{in-situ}} acceleration models struggle to produce the flat intensity profile as observed; how to broaden the CR distribution near the bubble surface but simultaneously match the multi-wavelength spectra will be a focus of future work. We also stress that because the observed bubbles are large and elongated, and  there is a significant variation in the properties within the Galactic halo (e.g., density, magnetic field strength, photon density of the ISRF), realistic 3D simulations with proper LOS projections will be crucial for direct comparisons to the data in order to examine the success of the models. 

Multi-messenger observational data will continue to provide critical constraints on the formation scenarios of the bubbles. X-ray and UV observations of the Galactic halo with increased number of sightlines and improved analysis techniques will offer better constraints on the thermal and kinematic properties of the halo gas. In particular, metallicity measurements enabled by future X-ray missions with high spectroscopic resolution (e.g., {\it XARM} \cite{Takahashi12} and {\it Athena} \cite{Nandra13}) may be used to distinguish between NSF and AGN events \cite{Inoue15}. Nonetheless, due to the difficulties mentioned in \S~\ref{sec:obs}, we encourage comparisons between theories and observations to be made in the data space in order to minimize the need for simplified assumptions. Observations ranging from GeV to PeV gamma rays (including {\it Fermi, HAWC, Cherenkov Telescope Array,} and {\it LHAASO}) as well as neutrinos (including {\it IceCube} and {\it ANTARES}) will provide improved constraints on the cutoff energy in the bubble spectrum, which is crucial for determining whether the bubbles are leptonic or hadronic in nature. Future observations in MeV gamma rays, together with {\it Fermi}, will also determine the existence of a spectral break at the threshold energy of pion production for the hadronic process at around a GeV. All of the above will bring new insights into the mysterious formation of the {\it Fermi} bubbles and our knowledge of feedback processes in our Milky Way galaxy and other galaxies.  
\vspace{6pt} 

\acknowledgments{H.Y.K.Y. thanks the organizing committee of the workshop "Three Elephants in the Gamma-ray Sky" for their invitations and hospitality in Garmisch-Partenkirchen, Germany, as well as the workshop participants, including Roland Crocker, Phillip Mertsch, Tracy Slatyer, Yoshiyuki Inoue, and Christopher Pfrommer for inspirational discussions. H.Y.K.Y. acknowledges support from NSF grant AST 1713722 and NASA ATP (grant number NNX17AK70G). M.R. acknowledges support from NASA grant NASA ATP 12-ATP12-0017 and NSF grant AST 1715140. EGZ acknowledges support from NSF AST 1616037, from the WARF Foundation, and from the University of Chicago, where part of this work was completed.}

\authorcontributions{H.-Y.K.Y. wrote the draft of the paper; M.R. and E.G.Z. improved the manuscript by contributing text and providing insightful comments.} 

\conflictofinterests{The authors declare no conflict of interest.} 

\abbreviations{The following abbreviations are used in this manuscript:\\

\noindent GC: Galactic center\\
NSF: nuclear star formation\\
AGN: active galactic nucleus\\
CR: cosmic ray\\
S-PASS: S-band Polarization All Sky Survey\\
HAWC: High Altitude Water Cherenkov\\
CRp: cosmic ray protons\\
IC: inverse Compton\\
ISRF: interstellar radiation field\\
CRe: cosmic ray electrons\\
3D: three dimensional\\
MHD: magnetohydrodynamics\\
LOS: line of sight\\
TDE: tidal disruption event}

\bibliographystyle{mdpi}


\bibliography{fb}

\end{document}